# INTEGRAL Cross-calibration Status: Crab observations between 3 keV and 1 MeV

**E. Jourdain**
*CESR/UT-CNRS*
*9 avenue du Colonel Roche, 31028 Toulouse cedex 04 - France*
*E-mail:* `jourdain@cesr.fr`

**D. Götz**
*CEA-Service d'Astrophysique*
*Orme des Merisiers, Bat. 709, 91191 Gif sur Yvette - France*
*E-mail:* `dgotz@cea.fr`

**N. J. Westergaard**
*DTU Space - National Space Institute*
*Technical University of Denmark, Juliane Maries Vej 30, DK 2100 Copenhagen O*
*E- mail:* `njw@space.dtu.dk`

**L. Natalucci**
*INAF-Istituto di Astrofisica Spaziale e Fisica Cosmica*
*Area Ricerca CNR/Roma 2 - Tor Vergata, Via del Fosso del Cavaliere, 100, 00133 Roma - Italy*
*E-mail:* `Lorenzo.Natalucci@iasf-roma.inaf.it`

**J. P. Roques**
*CESR/UT-CNRS*
*9 avenue du Colonel Roche, 31028 Toulouse cedex 04 - France*
*E-mail:* `roques@cesr.fr`

*On behalf of the instrument teams*

We present results of Crab observations by the INTEGRAL instruments. A simultaneous fit allows us to demonstrate that INTEGRAL provides reliable spectra over its wide energy range.

## 1. Introduction

This report presents the status of INTEGRAL instrument calibration after the OSA-7.0 release. We first summarize the status of each instrument, then a common fit will illustrate the inter-calibration picture.

## 2. Instrument calibrations:

### 2.1 JEM-X

The results reported here are based on the OSA-7.0 release with standard spectral extraction. Work is in progress to include the modelling of the gain effects into the next release of OSA.

**Orbit 300 data: 15 ks of useful duration** (ARF derived from earlier observations)

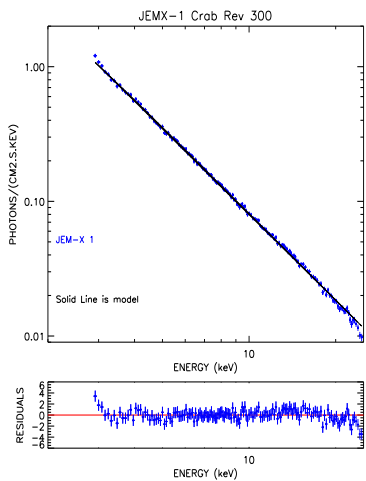 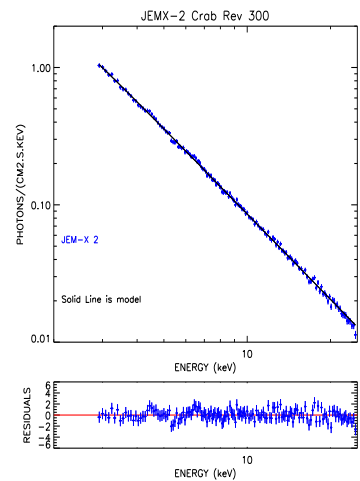

 JMX1                                          JMX2
Photon Index = 2.15 ± 0.05           Photon Index = 2.08 ± 0.05
Flux @ 1 keV = 11.4 ± 0.1 ph / cm$^2$ s keV     Flux @ 1 keV = 10.3 ± 0.1 ph / cm$^2$ s keV

$N_H = 0.361 \times 10^{22}$ cm$^{-2}$ has been frozen
For both instruments, Flux[2 – 10 keV] = $2.28 \times 10^{-8}$ erg / cm$^2$ s

## 2.2 **IBIS**

Latest results from IBIS/ISGRI calibrated mass model (Power Law fits):

**Orbit 300 (16 ks of useful duration):**
Photon Index = $2.12 \pm 0.03$
Flux @ 100 keV = $6.5 \times 10^{-4}$ ph / cm$^2$ s keV

**Orbit 605 (60 ks of useful duration):**
Photon Index = $2.13 \pm 0.03$
Flux @ 100 keV = $6.2 \times 10^{-4}$ ph / cm$^2$ s keV

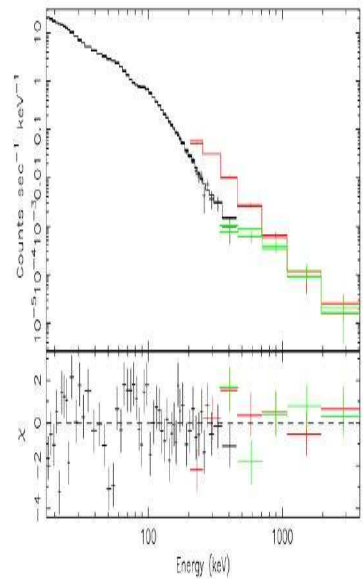

Crab spectrum measured by the IBIS instrument. ISGRI data (black) are collected during rev.300 (net time: 16 ks) and PICSIT data (red: single events; green: double events) are from revs. 39-45 (net time: 563 ks). Spectra are obtained using OSA7.0.

PICSIT data are treated with an average systematic error of 5% while ISGRI data have been added 1% systematics to each channel.

The reduced chi-square is 1.73 (61 dof).

Results are compatible with SPI ones at a 2σ confidence level.

With relatively a good performance below 100 keV, IBIS/ISGRI can effectively measure Crab spectrum and flux.
Above 100 keV a spectral break is detected but its position is not accurately determined so we fix it at 100 keV. More work is needed to assess the response at high energies, and to get rid of "narrow band" residual systematics below 100 keV.

The ISGRI response included in OSA-7.0 is derived from MC simulations and corrected a posteriori to match the SPI Crab spectrum.
Work is on-going to get rid of this ad-hoc correction by modelling the time dependent ISGRI response, using data from proton counters.

## 2.3 <u>SPI</u>

Considering the remarkable stability of the instrument, we have chosen to build a total Crab spectrum with all data available with 17 detectors and $5 \times 5$ dithering patterns (See poster by Jourdain & Roques, these proceedings, for more details).

This results in 326 ks of useful duration, distributed over 7 revolutions, from September 2004 to September 2007.

When going toward higher energies (~ above 100 keV), we can see that the global shape presents a slight curvature that can be approximated by a break in the (low energy) power law, suggesting a broken power law model. However, in this framework, we are facing a strong degeneracy in the parameter determination that we have chosen to solve by fixing the energy of the break at 100 keV.

Data are adjusted from 23 keV to 1 MeV with 0% systematics.

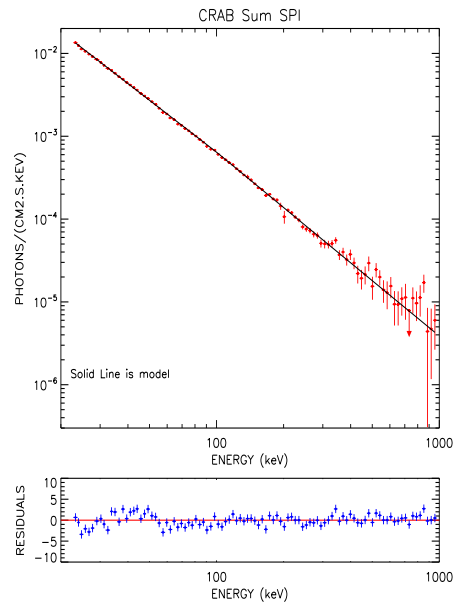

**Model :** *a broken power law with $E_{break}$ fixed to 100 keV*

  Photon Index 1 :      2.08 ± 0.01
  Break Energy  :    100.000   keV  frozen
  Photon Index 2 :      2.23 ± 0.03

Flux @ 100 keV:  $6.4 \times 10^{-4}$ ph / cm$^2$ s keV

Extrapolated flux @ 1 keV :
        9.3   ph / cm$^2$ s keV

A single power law is rejected:
An F-test relative to the broken power law excludes it with a probability of $2 \times 10^{-10}$.

.

## 3. The INTEGRAL Crab:

We then fit simultaneously JEM-X, ISGRI and SPI data with an absorbed broken power law with $E_{break}$ fixed to 100 keV and $N_H$ fixed to $0.361\ 10^{22}$ cm$^{-2}$.

**JEM-X 1 / 2** data have been used from 3 to 25 keV (15 ks in revolution 300, see Section 2.1).
**ISGRI** data have been used from 14 keV to 1 MeV (16 ks in revolution 300), but with OSA-7.0 standard response matrix (see Section 2.2).
**SPI** data have been used from 23 keV to 1 MeV (see Section 2.3).
     Systematics have been added at a level of 3% for JEM X-1 & 2 and 1% for ISGRI and SPI.

### *Results:*

*Model*: a broken power law with $E_{break}$ fixed to 100 keV
    plus an absorption with $N_H$ fixed to $0.361 \times 10^{22}$ cm$^{-2}$.

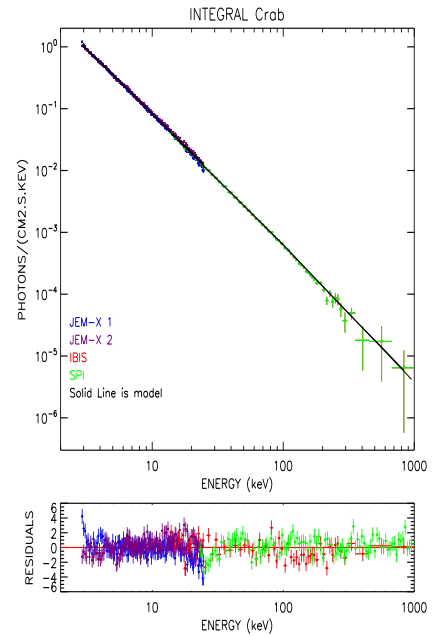

Photon Index 1 :    $2.105 \pm 0.3\ 10^{-2}$
Break Energy   :    100.000    keV (frozen)
Photon Index 2 :    $2.22 \pm 0.2\ 10^{-1}$

Flux @ 100 keV:   $6.3 \times 10^{-4}$ ph / cm$^2$ s keV

factor    SPI fixed to   1.0
factor    ISGRI   :    $0.99 \pm 0.2\ 10^{-2}$
factor    JEM X-1 :    $1.022 \pm 0.3\ 10^{-2}$
factor    JEM X-2 :    $1.06 \pm 0.3\ 10^{-2}$

Notes:
Residuals are in sigma units in all figures.
Statistical errors on best-fit parameters have been calculated with Xspec but are just indicative as complete ones depend on systematic level, binning, etc.

## 4. Conclusion

     The INTEGRAL instruments give consistent results on the Crab Nebula spectrum, with a global shape in agreement with observations from previous experiments.
Moreover, we have shown that for the ARFs and RMF provided with OSA-7.0, the cross-normalisation factors between SPI, ISGRI and JEM-X are compatible with 1.0 within reasonably small errors, and the shape of the combined spectrum is consistent with those for the single instruments in the overlapping energy windows.
This demonstrates that INTEGRAL can provide reliable spectra over its wide energy range.
     **This paper is mainly based on a document provided to the Integral User Group by instrument teams.**
         See     http://sigma-2.cesr.fr/spi/analysis/Cross-calibration.IUG.v2.pdf